\documentclass[prc,reprint,amsmath,amssymb,aps,showpacs,nofootinbib]{revtex4-1}
\usepackage{epsfig}
\usepackage{dcolumn}
\usepackage{bm}
\graphicspath{{figs/}} 
\begin{document}
\title{Measurement of the Branching Ratio for the Beta Decay of $^{14}$O}
\author{P.A.\ Voytas}
\email{Email contact: pvoytas@wittenberg.edu}
\affiliation{Physics Department, Wittenberg University, 
Springfield Ohio 45501, USA}
\author{E.A.\ George}
\affiliation{Physics Department, Wittenberg University, 
Springfield Ohio 45501, USA}
\author{G.W.\ Severin}
\altaffiliation{Present address: The Hevesy Laboratory, Center 
for Nuclear Technologies, Technical University of Denmark, 
Frederiksborgvej 399, 4000 Roskilde, Denmark.}
\affiliation{Physics Department, University of Wisconsin-Madison, Madison,
Wisconsin 53706, USA}
\author{L.\ Zhan}
\altaffiliation{Present address: Facebook Inc., 1 Hacker Way, Menlo Park, CA 94025}
\affiliation{Physics Department, University of Wisconsin-Madison, Madison,
Wisconsin 53706, USA}
\author{L.D.\ Knutson}
\affiliation{Physics Department, University of Wisconsin-Madison, Madison,
Wisconsin 53706, USA}
\date{\today}
\begin{abstract}
We present a new measurement of the branching ratio for the decay of 
$^{14}$O to the ground state of $^{14}$N.  The experimental result,
$\lambda _0/\lambda _{\rm total} = (4.934 \pm 0.040\kern 1pt{\rm (stat.)} 
\pm 0.061\kern 1pt{\rm (syst.)}) \times 10^{-3}$, is significantly
smaller than previous determinations of this quantity.
The new measurement allows an improved determination of
the partial halflife for the superallowed $0^+ \rightarrow 0^+$ 
Fermi decay to the $^{14}$N first excited state, which
impacts the determination of the $V_{ud}$ element of the CKM matrix.
With the new measurement in place, the corrected $^{14}$O ${\cal F} t $ value
is in good agreement with the average ${\cal F} t $ for other
superallowed $0^+ \rightarrow 0^+$ Fermi decays.
\end{abstract}
\maketitle

\section{Introduction}
\label{Intro}

The Cabibbo-Kobayashi-Maskawa (CKM) matrix parametrizes the extent to which
quark energy eigenstates are mixed in charge-changing weak decay processes.
For example, the weak interaction couples the $u$ quark 
to a mixture of $d$, $s$ and $b$ quarks, and the ``weak interaction 
eigenstate''
can be written in the form 
\begin{equation}
\vert d'\rangle = V_{ud}\vert d\rangle + V_{us}\vert s\rangle + 
V_{ub}\vert b\rangle,
\end{equation}
where $\vert d\rangle$, $\vert s\rangle$ and $\vert b\rangle$ 
are the quark mass eigenstates, and where $V_{ud}$, $V_{us}$ and $V_{ub}$ 
are elements of the CKM matrix.

Nuclear $\beta$ decay involves weak transitions between $u$ and $d$ quarks, and
one consequence is that decay rates are proportional to $\vert V_{ud}\vert^2$.
In fact, the value of $ V_{ud}$ is most accurately determined 
from an analysis of measured rates for superallowed Fermi decays, i.e., for
$0^+ \rightarrow 0^+$ transitions
between nuclear isobaric analog states.  
In a recent comprehensive analysis of the world data on decays of this kind,
 Hardy and Towner (HT) \cite{HT} 
 report the result $V_{ud} = 0.97417\pm 0.00021$.

One of the important isotopes in the analysis of Ref.\ \cite{HT} is
$^{14}$O.  The nucleus $^{14}$O has a $0^+$ ground state
and a half-life of
70.62\kern 2pt s \cite {thalf}, and decays by positron emission.
More than 99\% of the decays proceed by the superallowed 
Fermi transition to the 2.313 MeV $0^+$ first excited
state of $^{14}$N, while most of the remaining decays populate 
the $^{14}$N ground state.
The lifetime of $^{14}$O is well known \cite {thalf}, 
but determination of the  $ft$ value for the $0^+ \rightarrow 0^+$ transition 
requires knowledge of the ground state branching ratio, $R$.  

We are aware of three previous measurements of this quantity.
Sherr, et al.\ \cite{br1} report the value $R = (0.6\pm 0.1)\%$, while 
Frick, et al.\ \cite{br2} find $R = (0.65\pm 0.05)\%$.  Most recently, in 1966, 
Sidhu and Gerhart \cite{sidhu} obtained the result 
$R = (0.61\pm 0.01)\%$.  

All three measurements have been discussed and reanalyzed by
Towner and Hardy \cite{TH}, and based on that reanalysis HT have 
adopted  the value $R = (0.571\pm 0.068)\%$ for use in the analysis
of the superallowed $0^+ \rightarrow 0^+$ transitions.
We will comment further on the previous determinations of $R$
 in Sect.\ \ref{disc}.

In this paper we shall present a new measurement of the branching ratio, $R$.
When combined with the recent precise determination of the
$^{14}$O $Q$ value \cite{valverde}, the new measurement leads
to a significant reduction in the uncertainty of the 
$^{14}$O ${\cal F}t$ value.

\section{Description of the Experiment}
\label{Apparatus}

The measurements were carried out at the University of Wisconsin Nuclear
Physics Laboratory.  Many of the experimental details, including
a description of the apparatus, are given in a previous publication 
\cite{shape}, denoted in the following as GVSK.
Briefly, radioactive $^{14}$O is produced by bombarding a nitrogen gas target
with protons of about 8 MeV.  $^{14}$O produced in the target
cell is incorporated
into water, separated cryogenically from the nitrogen gas, and delivered
to a beta spectrometer.  There the sample is deposited
onto a cold, 13\kern 2pt ${\rm  \mu m}$ thick aluminum backing foil and
inserted into the spectrometer for counting.

The spectrometer \cite{betaspec} was constructed following 
the basic design principles of a ``Wu Spectrometer'' \cite {AL56}
with fields provided 
by superconducting magnets.
Positrons that pass through the spectrometer are detected in a nominally
5 mm thick lithium-drifted silicon [Si(Li)] detector.
The acceptance function of the spectrometer has a FWHM of about 2\%
and a peak solid angle of roughly 0.5 sr.  The centroid of the
acceptance function occurs at a momentum of approximately 248 keV/c per A 
of current, and the calibration is known to an accuracy of better than 
1 part in $10^4$ \cite {betaspec}.  

Measurements of the beta spectra for the ground  and excited state transitions
were obtained with different experimental procedures.  
The need for separate procedures is brought on 
by the presence of low energy positrons from
decay of $^{15}$O, produced by $^{15}{\rm N}(p,n)$, and from
$^{11}$C, produced by $^{14}{\rm N}(p,\alpha)$.  For the ground state 
measurements the counting rates are low, but the positron
energies are high and effects from the contaminant positrons are, for the
most part, irrelevant. On the other hand, the excited state measurements 
benefit from much higher rates, but counts from the contaminants 
need to be eliminated.

\subsection{Ground State Measurements}

Most of the relevant experimental details concerning the ground
state measurements are given in GVSK.  In that publication
we reported measurements of the shape of the ground state 
$\beta $ spectrum for positron kinetic energies ranging from 1.9 to 4.0 MeV.
The lowest energy is 90 keV above the endpoint of the excited state
decay, while the highest is close to the ground state endpoint energy,
4.12 MeV.

In the GVSK experiment,
the spectrum shape was determined by preparing
an $^{14}$O source and recording the number of 
detected positrons at several spectrometer currents in sequence.  The
process of source preparation followed by measurements 
at several currents, referred to as a cycle, 
was repeated many times to achieve the desired
statistical accuracy.  The measurements presented in GVSK determine ratios
of the ground state beta spectrum intensity at different currents, 
but do not fix the absolute normalization.

To obtain the branching ratio, we need to make a connection between 
measurements taken above and below the endpoint of the excited state 
transition.  The connection is made by using measurements at a 
spectrometer current of 8.8 A as a point of contact.
At 8.8 A, the centroid of the acceptance function corresponds
to a positron kinetic energy of 1.731 MeV. This energy is a bit
below the 1.809 MeV excited state endpoint, but still high enough so
that the results are not affected by positrons from $^{15}$O or $^{11}$C.

To supplement the data presented in GVSK, we collected measurements
for cycles in which the spectrometer current was moved between 8.8 and 11.0 A,
the latter being well above the excited state endpoint.
Typically, a source would be counted for 15 s at 8.8 A and then
for 53 s at 11.0 A.  For the next source we would count for 60 s
at 11.0 A and then 30 s at 8.8 A.  The process was
repeated for many such cycles.

\begin{figure}
\centerline{\includegraphics[width=80mm]{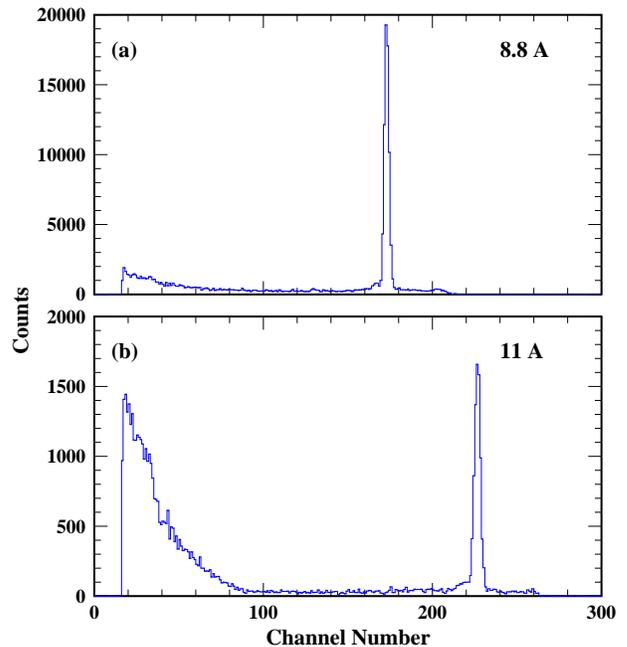}}
   \caption{(Color online) Si(Li) spectra obtained with the spectrometer current
set at  8.8\kern 2pt A and 11.0\kern 2pt A. These settings
are, respectively, just below and well above the endpoint of the excited state
transition.
The measured counts have been rescaled so that the two
spectra correspond to the same number of $^{14}$O decays.  Note the 
factor of 10 difference in scales for the two spectra.}
   \label{fig:spectra}
\end{figure}

Accumulated Si(Li) energy spectra for a typical run of this kind are shown
in Fig.\ \ref{fig:spectra}.  At each current there is a
primary peak corresponding to events in which the positron deposits its full
energy in the detector.  Counts above the main peak occur from processes
in which the positron energy pulse is supplemented with energy deposited by
one or both of the 511 keV annihilation $\gamma$ rays, while counts below
the primary peak but above channel 120 arise mainly from
events in which the positron backscatters out of the Si(Li) detector,
depositing only a portion of its kinetic energy.

Backgrounds from decay of $^{11}$C which has migrated to positions close
to the detector (see GVSK) and from 511 keV $\gamma$ rays originating
from inside the spectrometer are responsible for the increased
count rate below about channel 120.

The measurements were analyzed following the procedures outlined in GVSK.  
First, the raw counting rates are corrected for backgrounds.
We include room backgrounds, backgrounds
associated with the proton beam, and backgrounds arising from
2.3 MeV $\gamma$ rays emitted following decay of $^{14}$O
to the first excited state of $^{14}$N.
We also correct for  ``bad events'', which occur when positrons 
reach the Si(Li) detector after scattering from slit edges, the aluminum
backing foil, or other objects within the spectrometer.
The room and beam backgrounds are measured, while Monte Carlo simulations
are used to estimate the backgrounds from bad events and 
2.3 MeV $\gamma$ rays.

Let $S_R$ represent the measured number of counts within
some energy window for a given run and a given current.  If $B$ is
the associated background, then  
the corrected event sum $N_R(I)$ is given by
\begin{equation}
N_R(I) = (S_R-B)F_S/F_D.
\label{eq:NIx}
\end{equation}
Here $F_D$ is a decay factor defined as the fraction
of all $^{14}$O decays that occur while counting at a specific current, while 
$F_S$ corrects for good positron events that lie outside of the
summation window. We determined this latter quantity by using 
clean positron spectra from $^{66}$Ga decay in combination with Monte Carlo
simulations (see GVSK).
Basically, Eq.\ (\ref{eq:NIx}) corrects for
backgrounds and sub-threshold events, and extrapolates
the measured sums back to a common start time.

The quantities $N_R$ for a given run depend on the source activity
for that run and also on a quantity $\bar n(I)$, defined 
to be the beta spectrum intensity, 
$ n(p) \equiv {dn\over dp}$, integrated over the acceptance 
of the spectrometer.\footnote{We use $n(p)$ for the
total spectrum intensity, while $n_0(p)$ and 
$n_1(p)$ represent the contributions from
the ground and first excited state, respectively.}  
The activity factor cancels if we take ratios
of the corrected event sums, and consequently we are able to determine
the quantity
$\bar n(8.8 \kern 2pt {\rm A})/\bar n(11.0 \kern 2pt {\rm A})$.
The best-fit value of this ratio is found to be
$8.480 \pm 0.048$.


From this measurement we can determine the ratio of the spectrum
intensities, $n(p)$, at
the two currents. The acceptance width of the spectrometer scales
with the momentum,
and consequently $n(p)$ is obtained by
taking $ \bar n(I)/p$ and applying a small correction for the curvature
of the $\beta$ spectrum.  The result is 
\begin{equation}
{n(8.8\kern 2pt{\rm A})\over n(11.0\kern 2pt{\rm A})} = 
{n(2.183 \kern 2pt{\rm MeV/c})\over n(2.729\kern 2pt{\rm MeV/c})} = 
10.39 \pm 0.06,
\label {eq:nbar}
\end {equation}
where the quoted uncertainty includes statistics only.
With this result we can normalize the $\beta$ spectrum
of GVSK to the value at 8.8\kern 2pt A.  

In GVSK we assumed that the ground state $\beta$-spectrum 
intensity is of the form
%
%
\begin{equation}
n_0(p) = p^2\kern 2pt (E - E_0)^2\kern 2pt F(p,Z)\kern 2pt C(E),
\label{eq:theory}
\end{equation}
where $p$ ($E$) is the positron momentum (energy), $E_0$ is the endpoint 
energy, $F(p,Z)$ is a Fermi function, and $C(E)$ is a shape factor.
Our Fermi function is 
\begin{equation}
F(p,Z) = F_0(p,Z)\kern 2pt L_0^A \kern 2pt C_A\kern 2pt R_A\kern 2pt Q\kern 2pt
g(E,E_0),
\label {eq:FF}
\end{equation}
where $F_0$ is the usual Fermi function for a point charge nucleus
with lepton wave functions evaluated at the nuclear surface (see for example
Ref.\ \cite{wilk1}), $g(E,E_0)$ is a radiative correction factor
calculated following Sirlin \cite{sirlin}, $L_0^A$ and $C_A$ are finite size
corrections as given by Wilkinson \cite{wilk1}, and $R_A$ 
and $Q$ are corrections for recoil and screening, respectively, 
again calculated according to Wilkinson \cite{wilk2}.
Finally the shape factor is of the form
\begin {equation}
C(E) = k[1 + a'W + b'/W + c' (W-W_c)^2],
\label{eq:shapefunction_x}
\end {equation}
where $W$ is the positron total energy in units of its rest energy,
$a'$, $b'$ and $c'$ are constants, and $W_c$ is the $W$ value corresponding to
a positron kinetic energy of 2.75 MeV.  In GVSK, the shape parameters $a'$ and
$c'$ are determined by fitting measurements, while $b'$, which is
relatively unimportant, is fixed at its theoretical value.

\begin{figure}
\centerline{\includegraphics[width=80mm]{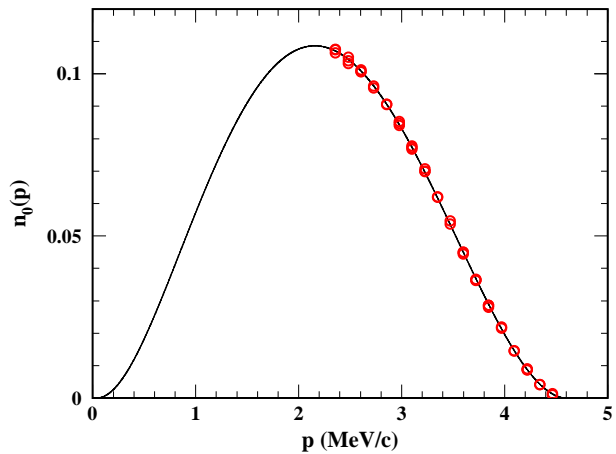}}
   \caption{(Color online) Measurements of the $^{14}$O $\beta$ spectrum intensity
for momentum values above the endpoint of the excited state 
transition.  The curve is from GVSK, with normalization fixed by 
the measurements reported here.  The open circles represent
a sample of the data used in GVSK to fix the shape of the $\beta$
spectrum.  
Either two or three points are plotted at each momentum value, since
the partial data sets have many overlap points.
The error bars, which are not plotted, are smaller than the symbols.
}
\label{fig:ground_state}
\end{figure}

The resulting ground state $\beta$-spectrum is shown by
the solid curve in Fig.\ \ref{fig:ground_state}.  The
curve is plotted on a scale set by the condition
$n(p) = 1$ at 8.8\kern 2pt A. This is accomplished
by using the 
the 8.8\kern 2pt A $\leftrightarrow$ 11.0\kern 2pt A measurements,
as summarized in Eq.\ (\ref{eq:nbar}),
to determine the constant $k$ in Eq.\ (\ref{eq:shapefunction_x}).
Effectively, what this does is to fix the $n(p)$
curve at $p=2.729 \kern 1.5pt {\rm MeV/c}$ to the value 0.0962.
Ultimately, we will use this calculated curve to determine the branching
ratio, and consequently it is important to know that 
 the spectrum shape is well determined.

The data points in Fig.\ \ref{fig:ground_state} represent a
sample of the measurements that were used in GVSK to determine
the $\beta$-spectrum shape, and are included to illustrate
the quality of the data set.  
To generate the points we collect data from all runs in which
measurements were taken at a particular sequence of currents, 
for example 10.0, 12.0, 14.0 and 16.0 A.
 Following procedures outlined
in GVSK, we fit this subset of measurements with a theoretical curve
of the form Eq.\ (\ref{eq:theory}) treating $a'$ and $c'$ of 
Eq.\ (\ref{eq:shapefunction_x}) as free parameters.
Each curve is then normalized to $n(2.729 \kern 1.5pt {\rm MeV/c}) = 0.0962$,
which fixes $k$, 
and the data (which lack an absolute normalization) are 
rescaled to the newly determined curve. 
This analysis was carried through 
for 10 independent subsets of the data.  
The subsets chosen were the ones that determined the shape parameters 
with greatest statistical precision.
All of the resulting points are shown in the plot.

\subsection{Excited State Measurements}
 
We must now obtain analogous results for decay to the
first excited state of $^{14}$N.
As noted earlier, 
the main complication in this case is the presence of positrons from
decay of $^{11}$C and $^{15}$O.  We separate out the contribution
from $^{14}$O by exploiting the fact that the three isotopes have  
different half lives.

The procedure is as follows.  In a given run we measure the
ratio of $\bar n(I)$  values at 8.8\kern 2pt A and some lower current.
In these runs, a single source is prepared and moved into the counting
position.  With the proton beam
turned off and the spectrometer current set at 8.8\kern 2pt A,
we observe decay positrons for a period of typically 200\kern 2pt s.  We
then change the current to some value in the range 2-8\kern 2pt A and
count for typically 1000\kern 2pt s.  Because the counting rates are
much higher for
the excited state transition than for the ground state, a few runs of this
kind are sufficient to give good statistical accuracy.

To analyze the measurements for a given run,
we first separate the data into segments in which the spectrometer current 
was fixed at one value or the other.  For each segment we
produce a Si(Li) energy spectrum.
We then choose a window around the peak in
the Si(Li) spectrum, and generate a decay curve, which gives counts within
the window as a function of time in 1\kern 2pt s intervals.

\begin{figure}
\centerline{\includegraphics[width=80mm]{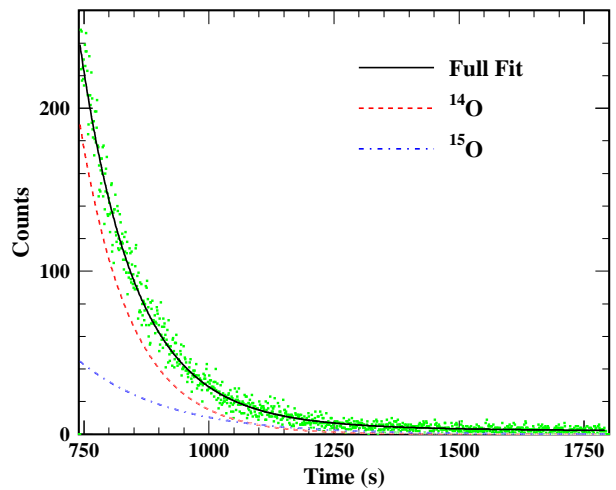}}
   \caption{(Color online) Sample decay curve for extracting the excited state spectrum by lifetime decomposition. 
Data shown are for a spectrometer current corresponding to a momentum of 1.985 MeV/c.
The points show the measured number of counts per 1 s interval, while the solid line is the fit obtained with the $A_i$'s of Eq. (\ref {eq:dk_curve}) treated as free parameters. 
The dashed and dot-dashed curves show the fit contributions from
$^{14}$O and   $^{15}$O.  The $^{11}$C contribution ranges from 4 to 2 counts/s.}
\label{fig:dk_curve}
\end{figure}

For the measurements at the lower current, we fit the decay curve data
with a three term expression,
\begin{equation}
N(t) = A_{1}\kern 1pt e^{-t/\tau _{1}} +A_{2}\kern 1pt e^{-t/\tau _{2}} +
A_{3}\kern 1pt e^{-t/\tau _{3}} ,
\label{eq:dk_curve}
\end{equation}
where $\tau _1$, $\tau _2$ and $\tau _3$ are the mean lifetimes for 
$^{14}$O, $^{15}$O and $^{11}$C, respectively, and where the $A$'s are the
fitting parameters.
Because the counting time is long compared to the mean lifetimes
of both $^{14}$O and $^{15}$O, the average Si(Li) rate at the
end of the counting interval is often less than 1/s.  Consequently,
we use Poisson statistics to optimize the fit to the decay curve measurements. 
A sample experimental decay curve and corresponding fit using Eq. (\ref{eq:dk_curve})
is shown in Fig. \ref{fig:dk_curve}.

\begin{figure}
\centerline{\includegraphics[width=80mm]{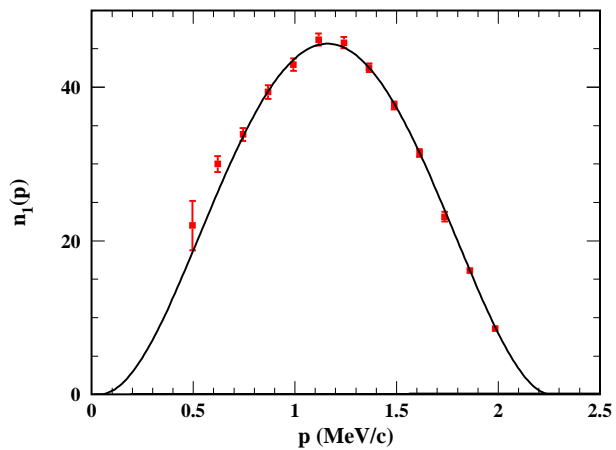}}
   \caption{(Color online) Measurements of the $^{14}$O $\beta$-spectrum intensity across
the region dominated by the transition to first
excited state of $^{14}$N.  The measured points were extracted from decay 
curves, and are plotted on a scale set by the condition 
$n(p) = 1$ at a current of 8.8\kern 2pt A.}
   \label{fig:exc_state}
\end{figure}

For the measurements at 8.8 A the acceptance of the spectrometer
is at or above the $^{15}$O and $^{11}$C endpoints.  Therefore, the
decay curve was fit with a formula that includes
an exponential term for $^{14}$O 
plus a constant to represent possible backgrounds.  In the initial fits
the background term was found to be statistically 
consistent with zero, and was subsequently 
fixed at zero for the final fits.

After obtaining the ratio of the $t=0$ $^{14}$O rates at the two currents,
we apply corrections for bad events, for backgrounds
from 2.3 MeV $\gamma$ rays, and for good positron events
that lie outside the Si(Li) summation window.
Finally we convert the resulting $\bar n$ ratio to a
ratio of $n(p)$ values.

The results are shown in Fig.\ \ref{fig:exc_state}.  
The plotted data points are the $n(p)$
values from which the small ground state contributions 
have been subtracted.  As in Fig.\ \ref{fig:ground_state}, the results are
normalized by the condition
$n(p) = 1$ at 8.8\kern 2pt A, and
the curve is once again a theoretical spectrum of the form
given in Eq.\ (\ref{eq:theory}), except that the shape factor $C(E)$
is now a constant.

For reasons to be explained later,
we only use the measurements for momenta above $p=0.9$\kern 1.5pt MeV/c 
in the determination of the branching ratio.  Thus, the
4 lowest-momentum points in Fig.\ \ref{fig:exc_state} are not employed.
All of the data points shown are well above the endpoint of the 
transition to the second excited state of $^{14}$N.

\section{Branching Ratio}
 
The full decay rate, $\lambda _i$, for either transition is obtained
by integrating the corresponding $\beta$-spectrum:
\begin{equation}
\lambda _i^{\beta} = \int n_i(p) \kern 2pt dp.
\end{equation}
From the best fit curves shown in Figs.\ \ref{fig:ground_state} and 
\ref{fig:exc_state} we obtain the ratio of the positron decay rates 
for the ground state ($\lambda _0^{\beta}$) and first excited state 
($\lambda _1^{\beta} $) transitions
\begin{equation}
{\lambda_{0}^{\beta}\over \lambda_{1}^{\beta}} = (4.965 \pm 0.040) \times 10^{-3},
\end{equation}
where the quoted uncertainty includes statistical errors only.
Taking into account decays to the second excited state of $^{14}$N, for 
which the average measured 
branching ratio is  $(0.545\pm 0.019)\times 10^{-3}$ \cite{HT},
and the relatively small rates for electron capture
into the ground and first excited state \cite {electron_capture},
we obtain a
ground state branching ratio 
\begin {equation}
R = {\lambda_{0}\over \lambda_{\rm total}} = (4.934 \pm 0.040) \times 10^{-3},
\end {equation}
where $\lambda_{0}$ and $ \lambda_{\rm total}$ include contributions
from both positron decay and electron capture.


\subsection{Systematic Errors}
\label{syst}

In Fig.\ \ref{fig:ground_state} we show the measured ground-state
$\beta $ spectrum.  These measurements cover only about half of the
full momentum range, so extraction of the
transition strength requires knowledge of the spectrum shape.
Our assumed $\beta $ spectrum uses a shape factor
parameterized in terms of the constants $a'$, $b'$ and $c'$ of
Eq.\ (\ref{eq:shapefunction_x}).

From GVSK
\begin{eqnarray}
\label{eq:result}
&& a' = -0.0290\pm 0.0008({\rm stat.})\pm 0.0006({\rm syst.}), \nonumber \\
&& b' = \kern 2.8mm 0.04\kern 3pt({\rm fixed}), \\
&& c' = \kern 2.8mm 0.0061 \pm 0.0010({\rm stat.})\pm 0.0005({\rm syst.}).\nonumber
\end{eqnarray}
As explained in GVSK, the quantity $W_c$ in  Eq. \ref{eq:shapefunction_x} was chosen so that the statistical uncertainties in $a'$ and $c'$ are uncorrelated.  
For the present analysis we take the uncertainties in $a'$ and $c'$ to be the linear sums
of the statistical and systematic errors listed in Eq.\ (\ref{eq:result}),
and $\delta b' = 0.02$.  The resulting systematic
uncertainties in $R$ are ${\bf \pm 0.009\times 10^{-3}}$ from $a'$, 
${\bf \pm 0.006\times 10^{-3}}$ from $b'$,
and ${\bf \pm 0.038\times 10^{-3}}$ from $c'$. 

In our experiment, the Si(Li) energy spectra  (see Fig.\ \ref{fig:spectra}) 
have large background rates below typically channel 120.
To determine the true event rate we sum the measured spectrum over
some range of channels (typically 146-240 at 8.8\kern 2pt A and
150-290 at 11.0\kern 2pt A) and then apply a correction for sub-threshold
events.  The procedure, which involves the use of experimental
$^{66}$Ga spectra and Monte Carlo simulations, is described in GVSK.
The uncertainty in the 11 A sub-threshold correction leads to an
uncertainty in the ground-state transition rate,
and thus to a possible systematic error in the branching ratio.
We estimate this systematic error to be ${\bf \pm 0.012\times 10^{-3}}$.

In our analysis we notice that the extracted branching ratio
changes by up to a few tenths of one percent as we move the
11 A lower summation threshold between channel 120 and channel 220 (see
Fig.\ \ref{fig:spectra}).  This variation is larger
than expected from uncertainties in the sub-threshold correction,
and to be on the safe side, we include an additional 
systematic uncertainty of ${\bf \pm 0.019\times 10^{-3}}$ to
cover this variation.

Sub-threshold corrections are required for the excited-state 
measurements as well as for the ground state.  
High-quality $^{66}$Ga reference
spectra are not available at these lower currents, although some
clean $^{14}$O spectra (obtained under different circumstances)
exist for currents between 4 and 8\kern 2pt A.
At these low currents, the sub-threshold counts
arise from events in which the positron backscatters out
of the Si(Li) detector without depositing its full energy.
Our experience is that our Monte Carlo simulations tend to over-predict the
number of such events \cite{shape}, and comparisons of the simulations
to the clean spectra
show an excess of about 7\% throughout the 4-8\kern 2pt A current range.
With that information in hand we calculate the sub-threshold corrections 
using simulations that have been scaled down by 7\% in the backscattering
region.  
To estimate the possible systematic error associated with
the correction, we take the scale factor to be $1.07\pm 0.07$.
The resulting systematic error in the branching ratio is 
${\bf \pm 0.032\times 10^{-3}}$.


Systematic errors in the 8.8 A sub-threshold correction cancel in the
determination of the branching ratio, since we use
the same summation windows and correction factors in 
the ground-state and excited-state analyses.

Contributions to the measured counting sums from background processes 
are small but not negligible.  The various background effects are discussed
in some detail in GVSK.
The largest effect in the present experiment is the presence of
counts within the 11\kern 2pt A summation window from 2.3 MeV $\gamma$ rays.
We use Monte Carlo simulations to estimate the number of such events, 
and these simulations are subject to possible systematic errors.
We estimate the resulting uncertainty in the branching ratio
to be ${\bf \pm 0.012\times 10^{-3}}$.

For the excited state transition, there are significant contributions
to the measured counting rates from events in which the positron
backscatters from the aluminum backing foil.  These and other ``bad events'' 
are estimated by Monte Carlo simulations and need to be subtracted.
The required corrections are large at low momenta.  For example,
for the two lowest currents shown in Fig.\ \ref{fig:exc_state}, the
bad event fractions are 17\% and 9\%.  
As noted earlier, we ignore the four lowest points in the determination
of the branching ratio, and the result is that the bad event fractions are
less than 2\% for the points retained.
The resulting systematic uncertainty in $R$ is 
${\bf \pm 0.006\times 10^{-3}}$.

To a good approximation, the spectrometer fields depend only on the magnet
current.  However, there are remnant fields from flux pinning in the
superconducting magnets, and these fields depend on the current history.
For the most part, the corrections are negligible except at 8.8 A where the
slope of $n(p)$ is large.  For the excited-state measurement we always 
approach 8.8 A from below, but the same is not true for the ground state
measurements.  We use a model \cite{betaspec} 
to estimate the pinning correction and
take the systematic uncertainty in $R$ to be
${\bf \pm 0.020\times 10^{-3}}$.

The net systematic error is obtained by summing the contributions listed
above in quadrature.  The result is 
\begin {equation}
\delta R = \pm 0.061\times 10^{-3}.
\end {equation}

A number of additional error sources were considered and found
to be insignificant.  These include uncertainties in the
ground and excited state $Q$-values, uncertainties in the
$^{14}$O and $^{15}$O half-lives, possible drifts in the spectrometer current,
the spectrometer calibration uncertainty,
and possible contributions
to the 8.8\kern 2pt A counting rate from $^{15}$O decay.

\subsection{Comparison to previous measurements}
\label{comp}

As we noted in Sec.\ \ref{Intro}, three previous 
measurements of the branching ratio have been reported. 

Sidhu and Gerhart (SG)
\cite {sidhu} obtained the value $R = (6.1 \pm 0.1)\times 10^{-3}$.  
A reanalysis of the
SG measurements that includes the effects of the various correction 
factors appearing 
in Eq.\ (\ref{eq:FF}) has been carried out by 
Towner and Hardy \cite{TH}.   They use $\beta$-decay shape factors derived from 
theoretical calculations that have been optimized to fit the SG data, and
obtain $R = (5.4 \pm 0.2)\times 10^{-3}$.  

We have also reanalyzed the SG data, and agree that the analysis
employed by SG results in a branching ratio that
is too high.  We use the spectrum shape reported
in GVSK and obtain $ R = 5.5\times 10^{-3}$.  
As seen previously in Ref.\ \cite{TH}, 
the fit to the SG data is not very good.  In our reanalysis the
total $\chi ^2$ is 76 for 11 data points.

Additionally, we believe that SG have failed to account for a non-trivial 
systematic uncertainty.
They determine the efficiency of their spectrometer by
measuring the intensity of the excited state transition at a kinetic
energy of 1.691 MeV.  At this point $n(p)$ has a rather large slope.
SG claim that their spectrometer calibration (momentum vs current) is known
to 1 part in $10^3$, and one finds that a $0.1\%$ change in $p$
translates into a change in $n(p)$ of more than 3\% at the calibration 
energy. Consequently, this effect, by itself, leads to an uncertainty of 
$\pm 0.2\times 10^{-3}$ in $R$.  
 
Taking into account the original $\pm 0.1 \times 10^{-3}$
error quoted by SG, the calibration uncertainty
noted above, and uncertainties associated with the shape of the 
ground state $\beta $ 
spectrum, we would quote
$R = (5.5 \pm 0.3)\times 10^{-3}$ for our reanalysis of the SG data.

Earlier determinations of the branching ratio were reported by
Sherr, et al.\ \cite{br1} and by
Frick, et al.\ \cite{br2}.  In both cases the authors determine $R$ from Kurie
plots which were analyzed in the allowed approximation, {\it i.e.}\ with a
Fermi function $F(p,Z) = F_0(p,Z)$ (see Eq.\ (\ref{eq:FF}))
and with a constant $C(E)$ shape factor.
  As pointed out by
Towner and Hardy \cite{TH}, the allowed approximation is not adequate 
for the ground state transition.

After describing their reanalysis of the measurements of 
Refs.\ \cite{br2,sidhu}, 
Towner and Hardy state ``The conclusion is clear.  For the Gamow-Teller 
branching ratio in $^{14}$O, determinations based on an allowed approximation
analysis of Kurie plots have to be increased by about 14\%.''
We do not agree with that conclusion.


Towner and Hardy draw their conclusion from analyses in 
which they determine the branching ratio by computing
\begin{equation}
R = {f_{\rm GT} \over f_{\rm F}}{X_{\rm GT}^2 \over X_{\rm F}^2},
\end{equation}
where
\begin{equation}
X=  {1\over n_i}\sum _{i=1}^{n_i} X_i,
\label{eq:x1}
\end{equation}
and
\begin{equation}
X_i=  {K(W_i)\over W_0 - W_i},
\label{eq:x2}
\end{equation}
and where $K(W_i)$ is either a data point from the Kurie plot for
analysis in the allowed approximation, or a Kurie point modified
by a shape correction factor for more accurate analyses.
In Eqs.\ ({\ref{eq:x1}) and ({\ref{eq:x2}), each point in the 
Kurie plot is given equal weight
in the determination of the transition strength $X$, independent of the
uncertainties in the $K(W_i)$ values and the subsequent $X_i$'s.

We extract branching ratios from the published data sets first in the 
allowed approximation and then with the
correction factors of Eqs.\ (\ref {eq:FF}) and
(\ref{eq:shapefunction_x}) (with the GVSK shape parameters) included.
The resulting branching ratios increase by 8.4\% for the data set of 
Ref.\ \cite{br2}, 11\% for the data set of Ref.\ \cite {sidhu},
and  6.3\% for our own data set, when the shape corrections are included.


Our reanalysis of the Ref.\ \cite{br2} data begins with the Kurie plots which
the authors show in Fig.\ 7.  Error bars are shown for 
three of the points plotted for  the ground state transition, 
and we take these to
be representative of the entire data set.  From the plotted $K(W_i)$
values we extract results for
$n(p)$ (along with uncertainties), and fit
the resulting data with the assumed $\beta $ spectrum 
by minimizing $\chi ^2$.  A similar procedure is used for the excited
state transition.  For the excited state the distribution of errors is
unimportant since the traditional allowed $\beta $ spectrum and the
modified spectrum of Eq.\ (\ref{eq:FF}) are very similar.
Our reanalysis gives $\lambda _0/\lambda _1 = 6.46\times 10^{-3} $ 
in the allowed approximation, in agreement with the result 
published in Ref.\ \cite{br2}.  When the shape corrections 
are included, we obtain
$\lambda _0/\lambda _1 = 7.0\times 10^{-3} $.

Following analogous procedures we have also reanalyzed the data of Sherr
et al.\ \cite{br1}.  Here only the ground state Kurie plot is shown, and
no visible error bars are displayed.  However, sample spectra are shown,
 which we integrate, using the sums to make error estimates.
(For this data set the measurements are confined to a relatively 
narrow momentum
range and consequently the deduced branching ratio is relatively 
insensitive to the assumed distribution of uncertainties.)
After extracting the $n(p)$ values we fit the 
data set with allowed and GVSK $\beta $ spectra,
and find that the GVSK spectrum gives a transition strength
which is 10\% larger than that of the allowed spectrum.
No plot is shown for the excited state transition, so we cannot extract
a spectrum integral for this transition.  
However, we know that the shifts in the extracted branching
ratios come almost entirely from the modification
of the ground state $\beta $ spectrum.  Therefore, 
our analysis, crude as it may be, suggests inflating the branching ratio
reported in Ref.\ \cite {br1}  from $6.0\times 10^{-3}$ 
to $6.6\times 10^{-3}$.

In summary, we quote $R = (5.5 \pm 0.3)\times 10^{-3}$ from Ref.\ \cite {sidhu},
$(6.6 \pm  1.0)\times 10^{-3}$ from Ref.\ \cite {br1},
$(7.0 \pm  0.5)\times 10^{-3}$ from Ref.\ \cite {br2},
and $(4.934 \pm 0.040 \pm 0.061 )\times 10^{-3}$ from the present work.

\section{Determination of the ${\cal F}t$ value}
\label{disc}

We now wish to see what effect the new measurement has on
the determination of the $^{14}$O ${\cal F}t$ value.
We will follow the procedures adopted by Hardy and Towner \cite {HT}.
First we take the error weighted average of the 4 branching ratio measurements
listed above, taking the net error in the present measurement
as the sum in quadrature of the systematic and statistical errors.  
The result is
$R = (5.013 \pm 0.070)\times 10^{-3}$.

The set of measurements deviate from the mean with a reduced $\chi ^2$
of 7.4.  In view of this, the uncertainty needs to be scaled up
by a factor $S$, which is taken to be the square root of
the reduced $\chi ^2$ recomputed
using only those measurements with errors less than $3\sqrt{N}$ times
the uncertainty in the average.  In the present case, this condition
eliminates Refs.\ \cite{br1} and \cite{br2} from the computation
of the scale factor.  We then obtain
$S = 1.95$ giving
\begin{equation}
R = (5.013 \pm 0.137)\times 10^{-3}.
\end{equation}
This is to be compared with the value
$R = (5.71\pm 0.68)\times 10^{-3}$ used previously in Ref.\ \cite{HT}.
With the new branching ratio, the uncertainty in the partial half
life for the $0^+$ decay is reduced from 50\kern 2pt ms to 15\kern 2pt ms.

In addition to the partial half-life, 
the value of ${\cal F}t$ also depends sensitively on the
$Q$-value of the decay.  The recent new measurement of this
quantity 
\cite{valverde} has reduced the uncertainty in $Q$ by nearly an order of 
magnitude.  When combined with previous measurements of $Q$
(see Ref.\ \cite{HT}), one obtains a weighted mean
of $Q_{EC} = 2831.564(0.027)$\kern 2pt keV.  
The corresponding value of the statistical weight factor, $f$,
is 42.8053(0.0027).  
This result was provided to us by Ian Towner \cite{f-value},
and was computed with codes that use exact solutions of
the Dirac equation.  The recent Towner-Hardy \cite{params} 
parametrization of the statistical weight factors gives the 
same result to the number of decimal places quoted.

Incorporating the updated value of $f$ along with the improved determination
of the partial half-life, 
we obtain a corrected ${\cal F}t$ value of
$3071.6 (1.9)$\kern 2pt s.  This is to be
 compared with $ 3071.4(3.2)$\kern 2pt s listed in Ref.\ \cite{HT},
and $ 3073.8(2.8)$\kern 2pt s 
from Ref.\ \cite{valverde}, where the new $Q$-value measurement was
incorporated for the first time.
Our new result is in good agreement with the average 
superallowed $0^+\rightarrow  0^+$ ${\cal F}t$ of $3072.27(0.62)$\kern 2pt s
reported in Ref. \cite {HT}.

\section{Conclusions}
\label{concl}
We have presented a new measurement of the $^{14}$O branching ratio which
is significantly more accurate than any previous determination.  
With the new result included, the uncertainty in
the weighted average of all measurements of this quantity 
is reduced by a factor of five.  Correspondingly,
the uncertainty in 
the partial half-life for the superallowed $0^+ \rightarrow 0^+$ 
Fermi decay to the $^{14}$N first excited state is reduced from 50\kern 2pt ms
to 15\kern 2pt ms.  Our new measurement combined
with the recent accurate determination of the $Q$-value
lead to an ${\cal F}t$ value whose uncertainty is reduced from 
3.2\kern 2pt s to 1.9\kern 2pt s.
At the present time, this uncertainty 
is completely dominated by correction factors that depend on
nuclear structure of the $A= 14$ $0^+$ states.

Our result for the $^{14}$O corrected ${\cal F}t$ value
is in good agreement with the average 
superallowed $0^+\rightarrow  0^+$ ${\cal F}t$ reported in Ref. \cite {HT}.
With the new $Q$-value measurement from Ref.\ \cite{valverde} and
our new determination of the branching ratio, $^{14}$O has the third lowest
${\cal F}t$ uncertainty of the ``traditional 14'' transitions which are
used in the determination of the CKM $V_{ud}$ matrix element.

The authors wish to thank Ian Towner and John Hardy for many helpful
comments during the course of the data analysis.  
We also thank Matthew Kowalski
for his assistance on the project, and the University of Wisconsin 
Center for High Throughput Computing for allowing access to their
computational facilities.
This work was supported in part by the National Science
Foundation under grants Nos. PHY-0855514 and PHY-0555649, and in part by
an allocation of time from the Ohio Supercomputer Center.

\end{document}